# A cross-sectional study of social inequities in medical crowdfunding campaigns in the United States

**AUTHORS:**

Nora Kenworthy[1]*

Zhihang Dong[2]

Anne Montgomery[3]

Emily Fuller[4]

Lauren Berliner[5]

[1] School of Nursing and Health Studies, University of Washington Bothell, Bothell, WA, USA

[2] Department of Statistics; Department of Sociology, University of Washington, Seattle, WA, USA

[3] Health Studies, Haverford College, Haverford, PA, USA

[4] Department of Philosophy, University of Washington, Seattle, WA, USA

[5] School of Interdisciplinary Arts and Sciences, University of Washington Bothell, Bothell, WA, USA

# Abstract

Americans are increasingly relying on crowdfunding to pay for the costs of healthcare. In medical crowdfunding (MCF), online platforms allow individuals to appeal to social networks to request donations for health and medical needs. Users are often told that success depends on how they organize and share their campaigns to increase social network engagement. However, experts have cautioned that MCF could exacerbate health and social disparities by amplifying the choices (and biases) of the crowd and leveraging these to determine who has access to financial support for healthcare. To date, research on potential axes of disparity in MCF, and their impacts on fundraising outcomes, has been limited. To answer these questions, this paper presents an exploratory cross-sectional study of a randomized sample of 637 MCF campaigns on the popular platform Gofundme, for which the race, gender, age, and relationships of campaigners and campaign recipients were categorized alongside campaign characteristics and outcomes. Using both descriptive and inferential statistics, the analysis examines race, gender, and age disparities in MCF use, and tests how these are associated with differential campaign outcomes. The results show systemic disparities in MCF use and outcomes: non-white users (and black women in particular) are under-represented; there is significant evidence of an additional digital care labor burden on women organizers of campaigns; and marginalized race and gender groups are associated with poorer fundraising outcomes. Outcomes are only minimally associated with campaign characteristics under users' control, such as photos, videos, and updates. These results corroborate widespread concerns how technology fuels health inequities, and about how crowdfunding may be creating an unequal and biased marketplace for those seeking financial support to access healthcare. Further research and better data access are needed to explore these dynamics more deeply and inform policy for this largely unregulated industry.

# 1. Introduction

Populations in the US and around the world are increasingly turning to online, donation-based crowdfunding platforms to solicit financial help for health care expenses.(1–4) On platforms such as the popular site GoFundMe, patients, family members, and supporters can build fundraising campaigns using text, photos, and video, and then easily distribute financial appeals to their extended social networks using emails, texts, or social media platforms. Medical campaigns make up more than a third of all fundraising efforts on sites like GoFundMe, raising more than $650 million a year.(5,6) GoFundMe has grown rapidly to control more than 90% of the donation-based crowdfunding market in the US: the total amount fundraised on the platform jumped from $1 billion to $4 billion between 2015 and 2017.(7,8) While crowdfunding is also used to fundraise for medical research and to support broader charitable causes, this paper will focus on its most common use – soliciting donations to cover personal medical and health-related costs, often referred to as medical crowdfunding.(9)

Medical crowdfunding (MCF) is a rapidly expanding and largely unregulated industry that is changing how Americans secure social and financial support in the midst of chronic and acute health crises. Many turn to MCF when other forms of healthcare coverage and social safety nets fail: a 2017 study of a randomized sample of US MCF campaigns found that they were disproportionately prevalent in states that had not accepted Medicaid expansion under the Affordable Care Act.(3) In the US, where 45% of adults are ranked as inadequately insured,(10) MCF can provide an important source of immediate material and social support. Despite common perceptions of crowdfunding as an easy way to gather capital, research has shown that only 10% of crowdfunding campaigns reach their financial goals, and many fall far short of success.(3,11) Numerous questions have been raised about how crowdfunding may exacerbate inequities by creating a competitive marketplace where only a few succeed. Many scholars have highlighted the likelihood that education, social class, demographic characteristics, attractiveness, type of medical condition, and size of social network could all impact the success of campaigns, and thus access to healthcare.(2,12–14) Yet with a few exceptions, empirical research on how these characteristics shape MCF access or success has been very limited.(3,15–17)

Despite widespread concerns among scholars about the ethical ramifications of MCF, and in particular its potential impacts on health disparities,(2,12–14,18,19) there has been very limited empirical research to help answer these questions and inform future policy regarding the industry. A 2019 paper by van Duynhoven and colleagues spatially assessed the geography of Canadian cancer campaigns, finding that they were most common in urban areas with higher levels of income, home ownership, and education.(15) Lukk and colleagues studied a relatively small sample of Canadian crowdfunding campaigns for individual education and health needs, and found that older adults, women, and "visible minorities" had poorer fundraising outcomes than other demographic groups.(16) Barcelos studied 410 crowdfunding campaigns for transition-related medical care for transgender people, and reported that the majority of campaigns were for white, transgender men seeking chest surgery, underscoring the importance of future intersectional analyses of crowdfunding inequities.(20) As Barcelos notes, these dynamics reflect the dominance of more privileged identities within the crowdfunding marketplace, and the likelihood that barriers to entry limit access to the marketplace as well as success within it.(20) In a mixed-methods analysis of 200 US MCF campaigns, Berliner and Kenworthy found that campaigners with the most severe and complex needs often struggled to create compelling campaigns that attracted donors.(3) These observations indicate that crowdfunding may be best suited to those who are higher in social hierarchies and experiencing simple, solvable problems; similarly, crowdfunding may present significant barriers to entry and market disadvantages to those who experience complex health and social issues.

To date, research on inequities in MCF has been hampered by several methodological challenges. The first challenge regards sampling: large crowdfunding platforms like GoFundMe will not release their data to the public, so researchers must rely the platforms' search and categorization tools to develop samples. This introduces a number of potential sources of bias, because platform algorithms are designed to prioritize specific content in making it visible to users. Typically campaigns that are more popular, that started more recently, and that are geographically proximate are more likely to appear earlier in search results. The second challenge is one of analysis. Because users are not asked by sites to list basic demographic information – including gender, age, race, or indicators of socioeconomic status (SES) – these indicators must be coded either by hand or machine learning. Both approaches, as discussed below, introduce potential errors and biases into the very data researchers are able to gather. Therefore, the ability to systematically measure demographic disparities in crowdfunding access or success has been limited. In response to these challenges, this paper presents the first known analysis of how demographic variables impact crowdfunding access and campaign outcomes using a randomized sample of medical campaigns in the US.

## 2. Potential sources of bias and disparity in crowdfunding

Extensive research has now shown that online social networks, platforms, algorithms, technologies, and even Silicon Valley corporations themselves are designed in ways that disproportionately disadvantage women, people of color, and other marginalized groups.(21–27) Ruha Benjamin has described this web of new technologies as a "subtle but no less hostile form of systemic bias," and one that often is disguised by common perceptions of technology as objective, open, and non-biased.(25) Research has documented systemic biases across many online technologies: search engine algorithms which reinforce racism and sexism online;(22) the collection, sale, and use of increasingly enormous troves of online data that disproportionately punish and surveil the poor and people of color and reinforce social hierarchies;(21,25,28–30) and social media sites and online algorithms that reduce racial empathy and reinforce and spread extremism and hate against marginalized groups (22,26,31,32). Given the complex and compounded axes of inequality and power that operate online, it is important to take a multi-dimensional approach to understanding both the hierarchies that produce technology, and the hierarchies that technology creates and reinforces. As a number of scholars note, the default identity that creates and structures the culture and systems of the modern internet is white, male, and capitalistic.(33–35)

Feminist scholars have also shown how the internet is changing (and often reinforcing) forms of labor that are often deeply feminized – including affective, relational, and emotional labor.(36–38) And while the internet has expanded the diversity of who does such labor (including, for example, a large number of men), it has not altered the dominant perceptions of such labor as feminine, or the ways in which such labor reinforces gendered care responsibilities both on- and offline.(36) Precarity in these new labor markets further accrues along lines of race as well as gender, with women of color and black women particularly burdened by these new labors, and in contexts where they are least protected.(38,39)

Additional dimensions of inequity require attention as technologies evolve, including along hierarchies of age, socioeconomic status, geography, and ability/disability. Because different technologies and platforms operate uniquely and cultivate different online cultures, disparities along these lines can be harder to predict or track, but are no less important. For example, many older adults struggle to acclimate to new tech infrastructures, while young people, who often master new technologies quickly, are also subject to exploitation within new online marketplaces such as YouTube due to their young age and lack of regulatory oversight.(40,41)

It is unlikely that crowdfunding would be immune to these dynamics, though research on the topic to date has been scarce. In fact, platforms like Gofundme are so integrated with other, more widely researched systems like Facebook and Google that it would be difficult for MCF to even separate itself technologically from sources of bias and discrimination on other platforms. Furthermore, Gofundme is impacted by the existing internet cultures, hierarchies and exclusions described above. These shape how people engage, assess each other, build relationships, accrue status, and leverage various forms of capital online. On similar platforms like Airbnb, such patterns of bias have manifested in measurable discrimination against non-white hosts and guests, as well as a self-discounting effect for non-white hosts.(42,43) Similar dynamics have been documented on entrepreneurial, rewards-based crowdfunding platforms like Kickstarter, with studies finding that African American men receive less funding and their projects are perceived as lower quality;(44) that minority producers face price discounting;(45) and that visually-identifiable race in Kickstarter campaigns impacts the probability of success.(46)

Disparities also extend to gender on these platforms, with Greenberg and Mollick finding that women are under-represented on these sites, receive less funding, and yet, with all other things being equal, would be expected to outperform men in their campaigns—a set of dynamics that also mimics workplace standards in more traditional industries.(47)

People of color, women, and others with marginalized identities experience at times acute discrimination and harassment online which has consequences ranging from emotional burnout to profound unsafety.(39,48) It can also lead people to avoid particular internet sites due to the social dynamics encountered there. Similar dynamics may also exist on MCF sites and impact who feels safe and welcomed on the platforms. For example, while Gofundme has been used as a space for organizing in support of racial justice, and to memorialize and bring awareness to racial injustice,(49) the platform has also been widely associated with prominent campaigns in support of nationalist or anti-black activism. This includes campaigns in support of armed militia to unlawfully police the US-Mexico border, and of white police officers involved in the deaths of unarmed black people. While Gofundme has later moved to shut down such campaigns, they continue to appear regularly on the site. Gofundme's past reluctance to shut down such campaigns was read by many as tacit endorsement. This, alongside observations that Gofundme directly profits by playing host to such causes, has led to campaigns such as the #dontfundhate campaign, protesting fundraising in support of Darren Wilson, the officer responsible for killing Michael Brown.(50) Incidents like these may lead to non-white communities feeling excluded from, under-represented by, or unsafe on, such platforms. Much more research is needed to document how people experience MCF platforms and their politics relative to their social identities and positions within powerful social hierarchies.

Finally, but no less importantly, crowdfunding platforms are embedded in broader social worlds where existing and historical disparities and biases may alter people's experiences of crowdfunding. Health system inequities and existing health disparities drive people toward crowdfunding; but people with less social, economic, or educational capital may find it harder to crowdfund, and thus may be less represented on the platform.(3,15) As described by Berliner and Kenworthy, successful crowdfunding requires multiple literacies and forms of expertise– digital, medical, social – as well as robust social networks with expendable capital.(3) MCF campaigns also appeal to, and reinforce, dominant social ideas of who is and is not deserving of charity. In the US in particular, these social mores are deeply rooted in histories of racial and gender oppression, whereby specific populations have been routinely classified as less deserving of social support, charitable assistance, and safety net programs—and have been publicly shamed for needing or asking for such assistance once it is available.(51–55) Existing social inequalities also compound the stigma of particular illnesses, increasing the likelihood that those with marginalized identities will be seen as to blame for their conditions.(56,57) All of this contributes to the likelihood of differential access and outcomes in medical crowdfunding campaigns. This study, which offers the first cross-sectional analysis of social inequities and their relationship to MCF campaign use and outcomes, is a much-needed first step in examining these dynamics.

Crowdfunding users are routinely told by experts and crowdfunding platforms themselves that they can improve their campaign outcomes by extensively engaging with networks and sharing compelling, multimedia narratives about their illness experience.(58–60) Yet given dynamics of inequity on platforms and the highly competitive market that MCF users face, it is likely that identities may play an equal, or larger, role in campaign set-up and outcomes. The approach below follows the conceptualization of Noble and Tynes,(34) recognizing the internet as "a system that *reflects,* and a site that *structures,* power and values" (p. 2, emph original). It is hypothesized that race and gender inequities are embedded into, and measurable at, several different parts of the crowdfunding process: 1) associated with use, measured as disparities in who is represented on crowdfunding sites; and 2) associated with outcomes, measured as disparities in campaign fundraising outcomes. Additional campaign characteristics, such as whether recipients are adults or children, and relationships between campaigners and recipients, are assessed for their association with campaign outcomes. As a counterpoint to this analysis, measures of social engagement between crowdfunders and donors / visitors are evaluated to assess for significant associations with campaign outcomes. As the two key domains of potential disparity (MCF use and outcomes) are analyzed below, a wide array of technological, social, and health-related factors of disparity that may contribute to these results are discussed and referenced. In another publication, Kenworthy elaborates a more comprehensive conceptual model of how these disparities may come into play in the crowdfunding process, to which readers are referred for further conceptual analysis of these dynamics.(61)

## 3. Methods

In order to conduct a more reliable assessment of who crowdfunds and how social factors are correlated with MCF campaign outcomes, a randomized sample of US MCF campaigns from GoFundMe, the most popular and widely used crowdfunding site in the US, was created and analyzed.(8) To create a randomized sample of campaigns, it was necessary to develop a robust and accurate sampling frame of medical campaigns on the site. To do this, a computer program was developed to search all medically-categorized campaigns on the site using US zip codes in July 2016. For each zip code searched, the site returned 500 campaigns closest to that zip code; this was done for every US zip code and duplicate campaigns were removed from the results. This process generated a list of 165,925 medical campaigns. From this list, a randomized sample of 822 campaigns was created for further analysis. Campaigns that were not primarily motivated by a medical cause or healthcare costs in the US or for a US resident were excluded. So, too, were campaigns for veterinary care, research efforts, medical volunteer work abroad, or fundraising on behalf of non-profit organizations. Out of 822 campaigns, 135 were excluded for failure to meet the inclusion criteria. An additional 47 campaigns were removed from Gofundme by the time of data analysis (July 2018), likely because campaigners had shut them down. In line with ethics guidelines for internet research,(62) these cases were not used since it was likely that the campaigners did not intend for their information to remain public and online. The remaining sample included 640 campaigns that met criteria and were still online by July 2018. Finally, 3 campaigns that had run for less than 30 days from when data were originally collected were removed from the sample so as not to skew results; ongoing qualitative research from this project suggests that most campaigns see the bulk of their donor activity in the first month, with steep declines in engagement and donations after that point in all but the rarest of cases.

    The University of Washington (UW) Human Subjects Division determined that Institutional Review Board approval was not required for this study because it used publicly available data and did not involve interactions or interventions. Nevertheless, additional efforts were undertaken to protect the data collected. All data was stored on password-protected, encrypted drives and de-identified prior to data analysis. Furthermore, the current publication presents only population-level statistics and not any individual or identifying data. All persons

working with identifiable data in the study analysis completed ethics training provided through the UW Office of Research.

Variables were created for a large number of characteristics extracted from each campaign page, including: outcome measures such as campaign goal, amount of money raised, number of donations, largest single donation, average donation, and percent of goal reached; campaign characteristics such as length (in days) and city and state where campaign was located; and social engagement measures such as shares, likes, comments, campaigner's number of Facebook friends, and the number of updates, photos, and videos. A number of additional demographic variables were identified and coded based on information from campaign sites and narratives. Campaigns were first coded for the perceived gender of both the MCF campaign organizer (the 'campaigner') and the intended recipient of the campaign funds (the 'recipient'). Sometimes these were the same person, but often someone was fundraising on behalf of another person. Gender was determined using three sources of data: first, pronouns and stated relationships (i.e., 'mother,' 'son,') in a campaign's text; second, user names, which were compared with the US Census baby names list by gender; and finally, user photos. If there was disagreement among these three sources of information, or if not enough information was available to make a conclusive judgement, the gender was marked as 'unknown.' Narratives were also coded for the relationship between the campaigner and recipient and whether the recipient was an adult (18 or older) or a child (under 18). These variables were only coded based on explicit, written information on the campaign page (for example, "My twelve-year old son…"); when information was not available these variables were coded as "unknown."

Finally, the perceived race of the campaign recipient was coded using three categories: 1) white; 2) black; 3) non-black person of color. While these are broad and imperfect categories that cannot capture the diversity or complexity of racial identification, experience, or discrimination in the US context, even simplistic racial binaries have been shown to map onto significant social and health inequities in the US.(63,64) There is no gold standard for assessing race online, although some strategies have been shown to be less objectionable than others: the use of facial recognition technologies in studies of social media is particularly problematic given embedded racial biases in those technologies and broader questions about how race can be defined from facial appearances alone.(65–67) For this project, perceptions of race by potential contributors to MCF campaigns are more relevant than how campaigners might identify themselves. Consequently, three raters from different racial backgrounds coded for race using visual and textual information drawn from campaign pages. Most often, there was not enough data on campaign pages to also assess the race of the campaign organizer, so only the race of campaign recipients was recorded. Using a fully crossed design, all raters assessed all campaigns. Intraclass correlation (ICC) was measured using the ICC two-way random test in SPSS.(68) This yielded an ICC score of .819, which is considered excellent for inter-examiner levels of agreement.(69)

The analysis aimed to answer two sets of questions: First, what are the demographics of MCF campaigners and recipients, and are there inequities in who is using, and providing the labor for, MCF and its intended beneficiaries? Second, are different demographic groups – including recipients with different gender, race, and age – associated with different campaign outcomes? In order to assess these questions, it was necessary to undertake two preliminary tasks. The first was identifying the best variable(s) for measuring campaign outcomes from the data available. While many people are concerned with 'success' in crowdfunding, this is difficult to measure given that all campaigners set different goals, and often have different expectations of what might constitute success. Thus, this project identifies fundraising 'outcomes' as a less value-laden measure of how campaigns performed. Possible variables for campaign outcomes included the total amount of money raised, the percentage of the goal reached, the average donation amount, and the number of donations. Measures of whether a goal was met, or the percentage of goal reached, are only valid if goal-setting is consistent among campaigners, which it is not. Many confounding factors, including the severity of illness and medical costs, may impact the goal that fundraisers set. Data from ongoing qualitative research also indicates that fundraisers set goals based on what they think they can reasonably ask for, not their actual financial need. Thus, using a percentage of the goal reached for campaigns may be more reflective of how campaigners set goals than of how successful their campaigns were. Goals themselves may impact the total amount of money campaigns raise by setting donor expectations and creating cut-off points for when campaigners may be compelled to stop campaigning – thus total amounts raised are also likely to be unreliable as measures of outcome.

Consequently, this analysis relies on two other dependent variables which represent the overall financial commitment a campaign generates: 1) the number of donations and 2) the average donation amount (calculated as total amount raised divided by the number of donations). These two measures do have some drawbacks, though they are less significant than those for other dependent variables. Average donation amount may not adequately capture popular campaigns' spread if they have many small donations, whereas the number of overall donations does not measure donation amounts. However, used side-by-side, these variables together provide a measure of both social network engagement with a campaign and the ways that donors are 'valuing' the campaign in the size of their donations.

The second preliminary task was assessing other campaign characteristics which might impact outcomes. In particular, campaigners are often told to engage extensively with their audience in order to boost the visibility and success of their campaigns. Thus, the analysis aimed to assess how social engagement between campaigners and potential donors – measured by updates, photos, videos, comments and hearts – corresponded with campaign outcomes. If these factors seemed significant to campaign success, they would need to be built into linear models as confounding variables. If they did not show significant correlation with campaign outcomes, this would lend credence to the hypothesis that demographic factors might play a larger role than others have predicted.

To carry out these analyses, data were compiled in SPSS and then further analyzed using R. Descriptive statistics on campaign characteristics and the demographics of campaigners and recipients were compiled. Descriptive statistics of campaigner's social engagement were also compiled, and pairwise Pearson correlations were used to identify significant covariation between these independent variables and campaign outcomes. Chi-square goodness of fit tests were used to assess how demographics of crowdfunding campaigners and recipients compare with the US population at large. Generalized linear models were then used to adjust for potential confounding variables, using the glm function in R. Linear and Poisson regression was then used to assess the relationship between various demographic characteristics and two dependent variables for campaign outcomes: the number of overall donations and the average donation amount.

# 4. Results
## 4.1. Campaign characteristics and social engagement
Medical crowdfunding campaigns in this sample show a wide variability in terms of campaign outcomes and social engagement. These basic descriptive results underscore how competitive the field of MCF is for crowdfunders, and how unequal MCF experiences and outcomes can be. As shown in Table 1, campaign goals – the total amount campaigners hoped to receive – ranged from $150 to $300,000 (mean 14753.06; SD 25278.14), and the total amount raised ranged from $0 to $45,000 (mean 3731.04; SD 5812.46). These figures show both the diverse purposes to which MCF is being put – from minor needs to major, high-cost interventions – and the wide range of outcomes on the platform. Tellingly, as other studies have found, (3,70) reaching financial goals is quite rare: only 9.2% of campaigns met their stated financial goal, and on average, campaigns reached less than half of their goal (mean 41.75%; SD 58.10%). There is repeated evidence of extreme disparities in terms of outcome, whether measured as the % of goal raised (min 0; max 864.50), the number of campaign donations (min 0; max 883), the average donation amount for each campaign (min 0; max 718.50), or the largest single donation, which ranged from $30 to $10,480. It is equally striking that some campaigns find such considerable financial success and engagement – even from single donors giving truly exceptional amounts of money – and that other campaigns find so little traction, with 2.3% netting no donations at all. Finally, many campaigns have been online for a long period of time (mean 458.14 days; SD 307.25), indicating not necessarily that they are active for this entire time, but, as is more likely, that they are left online after activity declines – possibly because campaigners forget about them, because taking campaigns down is more complex, or because pages archive a powerful moment in campaigners' lives. More research is needed into campaigner motivations and decision-making in this regard.

**Table 1. MCF Campaign Outcomes, Social Engagement, and Other Characteristics**

|  | Mean | Std. Dev | Min | Max |
|---|---|---|---|---|
| **Outcomes** | | | | |
| Campaign goal ($) | 14753.06 | 25278.14 | 150.00 | 300000.00 |
| Amount raised ($) | 3731.04 | 5812.46 | 0.00 | 45000.00 |
| % of goal raised | 41.75 | 58.10 | 0 | 864.50 |
| Donations (#) | 43.05 | 66.19 | 0 | 883 |
| Average donation ($) | 83.73 | 67.68 | 0 | 718.50 |
| Largest donation($) | 588.19 | 1008.97 | 0 | 10480.00 |
| **Social engagement** | | | | |
| Comments (#) | 1.12 | 2.01 | 0 | 16 |
| Updates (#) | 7.80 | 47.12 | 0 | 1200 |
| Photos (#) | 3.89 | 9.35 | 1 | 191 |
| Videos (#) | .07 | .45 | 0 | 6 |
| Shares (#) | 316.92 | 380.85 | 0 | 3600 |
| Hearts (#) | 41.42 | 64.10 | 0 | 919 |
| **Other characteristics** | | | | |
| Campaign length (days) | 458.14 | 307.25 | 30 | 1950 |

*Descriptive statistics of MCF campaigns from a randomized sample of 637 US medical crowdfunding campaigns on the site Gofundme.com.*

Alongside measures of financial success in campaigns, broader social engagement between campaigners and potential donors can be measured through media and updates posted by campaigners, and shares, likes (on GoFundMe these are 'hearts'), and comments by visitors. Here, too, there are wide disparities (Table 1): shares ranged from 0 to 3600 (mean 316.92; SD 380.85); hearts ranged from 0 to 919 (mean 41.42; SD 64.10); and comments ranged from 0 to 16 (mean 1.12; SD 2.01). It is worth noting, however, that these levels of engagement, even for the most successful campaigns in the sample, still fall far short of what is seen for truly viral campaigns by which campaigners may be inspired or even set expectations. As a point of comparison, one of the most successful and long-running medical campaigns on GoFundMe, "Saving Eliza," had, as of July 2019, 56,000 shares, 30,000 likes, and 260 comments.(71)

In addition to how potential donors engage with campaigns, campaigners engage with their publics through stories, updates (posts written after the campaign has launched), photos and videos. While campaigners are often told by sites like GoFundMe that such elements are essential to campaign success, there is little evidence to support this assertion, and some research shows that the use of photos, videos, and updates by users is more limited.(60) A study by Xu, for example, has shown that these elements are underused by campaigners, and similar results are shown here, particularly when it comes to videos.(72) Only 4.3% of campaigns included videos, for example, and while all campaigns had at least one photo, there was wide variation in how many photos and updates were posted (Table 1).

Covariates for social engagement, including the number of hearts (likes) and comments from visitors, as well as updates, photos, and videos from campaigners, were assessed for their association with campaign outcomes, measured as number of donations and average donation amount. As shown in supplemental figures S1 and S2, pairwise correlations of these covariates and the two outcome variables were used to test for marginal associations. Contrary to findings from previous studies, the data showed low correlations between updates, photos, and videos with these more sensitive outcome variables.(3,58,72) The number of hearts and comments were positively associated with the number of donations, though the effect was small. This may indicate that social media engagement elicits further donations, or that donors also engage in parallel activities of social media spread (i.e. sharing, liking) as a secondary contribution to campaigns. Overall, however, these covariates have a minimal or nonexistent association with outcomes, indicating that it is likely that other factors impact whether a campaign attracts donors, and how much donors give.

## 4.2. Who uses medical crowdfunding?

In order to answer the question of who uses MCF, both crowdfunding campaign organizers and intended recipients were coded for demographic characteristics whenever possible. In some cases, a person with a health issue or need builds a crowdfunding campaign for themselves, but more often, friends, family members, or neighbors start campaigns on behalf of patients who may be too incapacitated by illness to build their own campaigns or who feel too much shame about asking for financial help. Given that these social relationships are leveraged to attract donations from others in a social network, and often help to lend campaigns significant credibility, it was important to better understand campaigner identity as well as the relationship between campaigners and recipients. As shown in Table 2, only 20.41% (n=130) of campaigns were self-organized; 9.26% (n=59) of campaigns were organized by parents of the recipient; 22.29% (n=142) were organized by other immediate family members of recipients (siblings, spouses, grandparents, and children); 16.8% (n=107) were organized by friends or distant relatives; a handful were organized by unmarried partners (.47%, n=3); and in about a third of cases the relationship was not explicitly stated (30.77%).

**Table 2. Characteristics of crowdfunding campaigners and recipients**

|  | N | Percent |
| --- | --- | --- |
| **Campaigner-Recipient Relationship** | **637** | |
| *Self-fundraising* | 130 | 20.41% |
| *Parent* | 59 | 9.26% |
| *Immediate family member* | 142 | 22.29% |
| *Friend or distant relative* | 107 | 16.8% |
| *Unmarried partner* | 3 | .47% |
| *Unknown* | 196 | 30.77% |
| **Gender of recipient** | **637** | |
| *Man* | 300 | 47.10% |
| *Woman* | 318 | 49.92% |
| *Genderqueer* | 1 | .16% |
| *Unknown* | 18 | 2.83% |
| **Gender of campaigner – self-fundraising** | **127** | |
| *Man* | 41 | 32.28% |
| *Woman* | 85 | 66.93% |
| *Genderqueer* | 1 | .79% |
| *Unknown* | 0 | |
| **Gender of campaigner – fundraising for others** | **510** | |
| *Man* | 86 | 16.86% |
| *Woman* | 400 | 78.43% |
| *Genderqueer* | 0 | |
| *Unknown* | 24 | 4.71% |
| **Race of recipient** | **637** | |
| *White* | 495 | 77.71% |
| *Black* | 52 | 8.16% |
| *Non-black POC* | 66 | 10.36% |
| *Unknown* | 24 | 3.77% |
| **Age of recipient** | **637** | |
| *Adult* | 524 | 82.26% |
| *Child* | 113 | 17.74% |

Demographic characteristics of crowdfunding campaigners and recipients, including gender, race, and age, as well as relationship between campaigner and recipient, if specified in the campaign.

The gender of both recipients and campaigners was coded using techniques described in the methods section above. The results are shown in Table 2. Among campaign recipients, data showed a relatively equal gender balance, with 47.1% men (n=300), 49.92% women (n=318), and .16% genderqueer (n=1). Transgender identities were coded as such when self-identified in the data, but only 2 campaigns in the sample explicitly stated their transgender identity--both of them self-organized campaigns for transgender men. In recognition of trans people's self-identification of gender and the fact that many may pass as cisgender and/or not reveal their status as transgender due to safety or privacy concerns, all campaigns were coded according to participants' stated gender as men, women, or genderqueer. Given the popularity of crowdfunding for transgender health needs, including for gender confirmation treatments, it is worth highlighting that transgender campaigns were a very small population in the overall sample (n=2), comprising less than 1% of all cases.(20,73)

Despite the relative gender balance among male and female identified campaign recipients, there were acute gender imbalances among campaign organizers. Among those campaigning for themselves, about 67% (n=85) were women, and among those campaigning on behalf of others, nearly 80% (n=400) were women (see Table 2). As shown in Table 3, chi-square goodness of fit tests were conducted to compare the means for recipient and campaigner gender with the US population at large, drawing from the 2017 American Community Survey estimates. These tests confirmed that while there was no statistical difference between the gender balance of recipients and that of the population at large, there were highly statistically significant (p=.000) differences among campaigner gender, for those self-fundraising ($X^2$=13.993) and, particularly, for those fundraising for others ($X^2$=192.998). As discussed below, these gender imbalances indicate a strong gendered labor component to crowdfunding that has not yet been explored in the literature.

**Table 3. Means comparison of demographic characteristics between crowdfunding and US population at large**

|  | N | Percent | US Population at large (%)[a] | Comparison of means ($X^2$) | P value |
|---|---|---|---|---|---|
| **Gender of recipient** | | | | | |
| *Man* | 300 | 48.54% | 49.2% | .107 | .744 |
| *Woman* | 318 | 51.46% | 50.8% | | |
| **Gender of campaigner – self-fundraising** | | | | | |
| *Man* | 41 | 32.54% | 49.2% | 13.993 | .000*** |
| *Woman* | 85 | 67.46% | 50.8% | | |
| **Gender of campaigner – fundraising for others** | | | | | |
| *Man* | 86 | 17.70% | 49.2% | 192.998 | .000*** |
| *Woman* | 400 | 82.30% | 50.8% | | |
| **Race of recipient** | | | | | |
| *White* | 495 | 80.75% | 73% | 18.980 | .000*** |
| *Black* | 52 | 8.48% | 12.7% | | |
| *Non-black person of color* | 66 | 10.77% | 14.3% | | |
| **Age of recipient** | | | | | |
| *Adult (18+)* | 524 | 82.26% | 77.1% | 9.608 | .002** |
| *Child (under 18)* | 113 | 17.74% | 22.9% | | |

Means comparison between demographics of crowdfunding campaigners and recipients and US population at large. Unknown and genderqueer cases removed for the purposes of this analysis.
***=<.001; **=<.01; *=<.05
[a]Source for US population data: US Census Bureau, American Community Survey(74).

Using similar methods, the race of recipients was also examined, using 3 broad categories for perceived racial identity: 1) white; 2) black; 3) non-black person of color (POC). White people accounted for more than 75% of campaign recipients overall (77.71%, n=495), while black people represented 8% of recipients (n=52), and non-black POC represented 10% (n=66) (Table 2). As can be seen in Table 3, these proportions also show significant ($X^2$=18.980, p=.000) differences from the US population at large, with non-white groups significantly under-represented and whites over-represented across the board. Racial disparities are particularly notable here given research showing that racial minorities, particularly African Americans, are more likely to be uninsured;(75) have higher rates of chronic disease, premature death, and many types of injury;(63) and carry more medically-related debt(76) – factors which would increase the likelihood of conditions and financial needs which might cause people to crowdfund.

Given considerable evidence that technological disparities accrue most significantly among women of color,(34) it was important to use an intersectional frame to investigate gender and race representation together. Table 3 shows a breakdown of MCF users (n=554) by gender and race, with those with unknown race or gender removed. While white women are over-represented, under-representation is particularly acute for black women, who make up less than 7% of women in the sample. This resonates with numerous accounts of black women's marginality in online spaces, and is particularly striking given the health disparities and medical debts they face.

**Table 4: Cross-Tabular Description of Recipient Race and Gender (n=554)**

|  |  | Man | Woman | Genderqueer |
|---|---|---|---|---|
| **White** | Count | 210 | 237 | 0 |
|  | Expected count[a] | 215 | 219 | n/a |
|  | % within race | (77.49%) | (84.04%) |  |
| **Black** | Count | 27 | 19 | 1 |
|  | Expected count[a] | 34 | 38 | n/a |
|  | % within race | (9.96%) | (6.74%) |  |
| **Non-Black POC** | Count | 34 | 26 | 0 |
|  | Expected count[a] | 23 | 25 | n/a |
|  | % within race | (12.55%) | (9.22%) |  |
| **Total** |  | 271 | 282 | 1 |

Cross-tabular descriptive statistics of MCF recipients by race and gender, with expected counts based on US population estimates by race and gender.
[a]Source for US population data: US Census Bureau, American Community Survey(74).

Finally, the proportions of recipients who were adults or children were assessed (Tables 2 and 3). The majority of recipients (82.26%, n=524) were adults, a statistically larger proportion than the US population at large ($X^2$=9.608, p=.002). While there have been many high-profile crowdfunding campaigns for sick children, and children are often perceived as particularly deserving cases on the site, it is notable that they make up a smaller proportion of campaigns than might be expected.

## 4.3. Disparities in crowdfunding success

In addition to assessing under-representation of specific groups and the over-representation of female labor in MCF campaigns, disparities in campaign outcomes among different groups can also be measured. Two key measures of campaign outcome – the number of donations and the average donation amount—were assessed. The primary aim was to measure the associations between race, gender, and age of campaign recipients and campaign outcomes. The relationship between campaigner and recipient was also included, because of impact this can have on the perceived credibility, and thus potential outcomes, of campaigns.

To explore these relationships, a linear regression analysis of the effect of recipients' gender, race, age, and campaigner-recipient relationship on the average donation amount was conducted (Table 5). The relationship between campaigner and recipient did not have any significant association, nor did the gender of the recipient. However, significant effects were seen for both race and age factors. Being black was associated with a recipient receiving on average about $22 less per donation (-22.223, p=.030); while being a non-black person of color was associated with about $12 less per donation, this finding was not statistically significant (p=.178). Surprisingly, being a child was also associated with a lower donation amount, a loss of about $18 per donation (p=.022).

**Table 5. Linear Regression of Demographic Effects on Average Donation Amount**

|  | Estimate | 95% Confidence Interval | P-value |
|---|---|---|---|
| (Intercept) | 66.494 | (-22.131, 155.119) | 0.142 |
| **Relationship to Recipient** | | | |
| Self (Reference) | | | |
| Parent | 13.086 | (-10.148, 36.320) | 0.270 |
| Immediate Family Member | 7.151 | (-8.761, 23.063) | 0.379 |
| Unmarried Partner | 14.864 | (-60.774, 90.502) | 0.700 |
| Friend or Distant Relative | 13.03 | (-4.143, 30.204) | 0.139 |
| Unknown Relationship | 6.264 | (-9.003, 21.530) | 0.422 |
| **Gender** | | | |
| Male (Reference) | | | |
| Female | -2.738 | (-13.310, 7.833) | 0.612 |
| Genderqueer | 27.074 | (-103.654, 157.802) | 0.685 |
| **Race** | | | |
| White (Reference) | | | |
| Black | -22.223 | (-42.266, -2.179) | 0.030* |
| Non-Black POC | -12.1 | (-29.693, 5.493) | 0.178 |
| **Age of Recipient** | | | |
| Adult (Reference) | | | |
| Child | -18.124 | (-33.655, -2.593) | 0.022* |
| Log of Number of Residents in State | 1.287 | (-4.155, 6.730) | 0.643 |

Linear regression analyzing the relationship between gender, age, race, and campaigner-recipient relationship and average donation amount to MCF campaigns. ***=<.001; **=<.01; *=<.05

Second, a Poisson regression analysis of the relationship between recipients' gender, race, age, and campaigner-recipient relationship on the number of donations to the campaign was conducted (Table 6). Non-white recipients were likely to receive significantly fewer donations. Those for whom gender could not be determined given the available information were also associated with fewer donations, though this may be a reflection of the general paucity of information or unclear narratives in those campaigns. Similar to the findings for gender and average donation above, there was a small (but significant) difference between men and women in terms of the number of donations received, with women earning fewer donations. By contrast, child recipients received significantly more donations, as did those whose immediate family members, parents, or friends were campaigning on their behalf. These findings indicate that children attract a smaller average donation but more donations overall. Interpretations of this seeming paradox are discussed below.

**Table 6. Poisson Regression of Demographic Effects on Number of Donations**

|  | Estimate | 95% Confidence Interval | P-value |
|---|---|---|---|
| (Intercept) | 3.358 | (3.321, 3.395) | <0.001*** |
| **Relationship to Recipient** | | | |
| Self (Reference) | | | |
| Parent | 0.341 | (0.285, 0.396) | <0.001*** |
| Immediate Family Member | 0.580 | (0.540, 0.647) | <0.001*** |
| Unmarried Partner | 0.098 | (-0.107, 0.302) | 0.350 |
| Friend or Distant Relative | 0.375 | (0.331, 0.420) | <0.001*** |
| **Gender** | | | |
| Male (Reference) | | | |
| Female | -0.027 | (-0.050, -0.003) | 0.026* |
| Genderqueer | -1.508 | (-2.310, -0.706) | <0.001*** |
| Unknown | -0.237 | (-0.316, -0.158) | <0.001*** |
| **Age of Recipient** | | | |
| Adult (Reference) | | | |
| Child | 0.280 | (0.248, 0.311) | <0.001** |
| **Race** | | | |
| White (Reference) | | | |
| Black | -0.058 | (-0.104, -0.012) | 0.013* |
| Non-Black POC | -0.179 | (-0.222, -0.135) | <0.001** |

Poisson regression analyzing the relationship between gender, age, race, and campaigner-recipient relationship and number of donations to MCF campaigns. ***=<.001; **=<.01; *=<.05

# 5. Discussion

Proponents of crowdfunding have repeatedly emphasized that it is a way of using social media to democratize charity and philanthropy, expanding who has access to charitable giving and who can participate in it as donors.(77,78) By contrast, this research indicates that the crowd dynamics brought to these platforms are not the dynamics of a social democracy – wherein all citizens have shared, equal rights and protections – but of an aristocratic oligarchy, where an elite few succeed, and the majority of users struggle, facing multiple scales of hierarchy and inequity. Data on the demographics of crowdfunding users indicates that there are significant inequities in representation which may reflect issues in the accessibility of the platform or sociopolitical dynamics that make people with various marginalized social identities feel unwelcome or unlikely to succeed. Given existing evidence of rampant disparity, discrimination, and bias across internet spaces, technologies, and social media platforms, it is possible that dynamics carry over from other internet spaces or are forged anew on this site. In the case of crowdfunding, technological disparities are likely compounded by social biases which users bring to the platform, including expectations of who can and should be crowdfunding, and whose cases are the most trustworthy and deserving.(3,16,73) The impacts of these biases can be seen in the disparities in campaign outcomes by race, in particular, but also by gender and age.

      Gender and race disparities in terms of who uses MCF constitute a base layer of inequities in terms of representation on the site. The disproportionate number of crowdfunding recipients who are white likely reflects both technological and sociopolitical disparities which might deter non-white users, particularly black women, from appealing for charity using these platforms. These disparities are even more striking when considered in light of the disproportionately large numbers of people of color, particularly African Americans, who are sicker, less insured, and more medically indebted. Disparities in representation are compounded by disparities in outcome – evidence here that shows non-white users are likely to raise less money on the platform, measured by both the number of donations and the average donation amount. Alongside limited but compelling research showing evidence of systemic racial biases on platforms like Airbnb and Kickstarter,(43–46) there is good reason to suspect that charitable crowdfunding platforms offer new spaces for existing racial biases to be enacted, and potentially play their own role in exacerbating these biases by exposing charitable appeals to a broader audience or subjecting campaigns to their own internal algorithms which might be biased against particular kinds of users.

      Alongside these racial inequities are unequal gender dynamics and disparities between cisgender and non-cisgender users. It is notable that there are not significant differences in campaign outcome between men and women, except for in number of donations. However, these outcomes may mask a number of social dynamics and biases which are influencing MCF – such as gendered and racialized social mores about who should be entitled to social assistance (typically in the US this has been women and children); who is financially capable and trustworthy; who is more likely to be at fault for being sick; and who faces more shame for asking for financial assistance.(54,55) More overt disparities are at work among non-cisgender crowdfunding users, with very few overall represented on the site (despite research showing that crowdfunding is a popular option for fundraising for transgender medical care); and with much poorer outcomes among non-binary users, though the n of 1 is very small. These findings resonate with those of Barcelos, who found that transgender crowdfunding campaigns raise on average a much smaller amount of money than has been found in research on general medical campaigns.(20)

      The stark gender imbalances among campaign organizers represents a new frontier of gendered care labor which is being undertaken online and across digital social networks. Whether online or in person, gendered care labor often comes at the expense of self-care and the maintenance of one's own social network ties, which may be represented in the disproportionate number of women who are self-campaigning as well. Data here confirms that the vast majority of campaigns do not go viral, and thus the 'crowd' with whom crowdfunders are interacting is often an intimate, densely connected social assemblage. This may increase the gendered aspects of the labor involved in campaigns, as much of it involves managing intimate relationships among family, friends, neighbors, and co-workers, and repeatedly interacting with this network both collectively and individually over time as the campaign unfolds. This represents a new domain of gendered, digital care labor, which underscores how both digital labor and care labor are deeply

feminized within contemporary social and economic realms. This finding resonates with ample work in digital media studies showing how relational and affective labor is crucial to many practices and economies online, from the work of musicians engaging with fans, to the management of community forums, to the liking of status updates on Facebook.(36,37) Thus, crowdfunding platforms represent an amalgamation of various forms of often (unfairly) feminized labor, including work to: sustain relationships; catalogue, remember and manage exchanges within gift economies; keep information flowing among friends and family; manage emotions (of self and others); capture and share images; and foster and sustain empathetic connections.

Lastly, these results suggest interesting though unexpected dynamics of inequity between children and adults on crowdfunding platforms. The general under-representation of children within this sample contradicts the visibility of campaigns for children within the popular press and imagination, which may be driven by rare but highly viral campaigns such as "Saving Eliza".(71) Several other factors might contribute to the under-representation of children, including the reticence of parents to post private information about their children online, the relatively larger number of social and charitable support programs available to sick children, or the fact that children tend to face fewer hospitalizations and chronic illnesses than adults.(79,80) A seeming paradox in campaign outcomes emerges: children's campaigns attract more donations, but of lower average amounts. This may reflect a dynamic whereby children's' campaigns spread more widely through social media, perhaps because they elicit broad sympathy, but attract lower average donations as the donors contributing to their campaigns may not be as closely related to the family and thus willing to give less money. Thus, sympathy may not be synonymous with success on crowdfunding platforms – a dynamic that merits much more intensive analysis, particularly in terms of what motivates donors and how they decide what to give to specific campaigns.

## 6. Conclusion

This exploratory research presents the first known analysis of how basic demographic characteristics such as age, race, and gender are represented in MCF, and how these characteristics are correlated with crowdfunding outcomes. The data point to systemic disparities in terms of crowdfunding use and outcomes. It highlights how inequities can affect crowdfunding in different ways, such as the significant burden of digital care labor that falls on women, even while women and men have relatively similar campaign outcomes. While the study sample size was too small to explore it thoroughly, the data also point to the likelihood of intersectional dynamics by which race, gender, age, and other disparities compound one another within the crowdfunding environment. Overall, this paper provides essential evidence that crowdfunding is playing host to, and potentially exacerbating, social biases related to perceived deservingness; however, much more research is needed to better understand how these disparities are created, and the social and technological mechanisms through which they are sustained and compounded.

Data presented here indicate that existing technical and sociopolitical disparities act as barriers to entry for some people *and* affect the success of certain campaigns once they are set up. What this means is that people from marginalized groups may face two sets of inequities, both of which produce disparities in MCF, and which, taken together, compound each other. It is also likely that there is a feedback effect at work: if a person sees others in their demographic who are not successful in their MCF campaigns, they are probably less likely to turn to crowdfunding themselves when a need arises; conversely, if one sees others like them succeeding at crowdfunding, they may be more likely to use it themselves. Thus, outcomes among others in a social class or group can influence future use and success. Much more research is needed, however, to explore these dynamics in more detail, including the affective and personal dimensions of what it means to inhabit MCF spaces while coming from non-dominant identities and backgrounds.

There are several limitations to this study. The first (and most prominent) relates to the data researchers are able to access from Gofundme, which does not share data publicly. It is likely that Gofundme collects much more robust information on users' demographics and geographic locations, but without access to this data, researchers must rely on coding methods like those described above, or facial recognition technologies, which present a host of methodological and ethical issues, particularly when used with racially diverse populations. The demographics that can be assessed from the data available are limited. A more nuanced measure of age would strengthen the analysis considerably and enable assessments of whether dynamics like ageism are impacting campaign outcomes. Also notably absent from the data are measures of socioeconomic status and class, which are important variables on which research has been limited.(81) Future research exploring how different medical conditions impact campaign outcomes will be important as well. A larger data sample overall would also enable a more robust intersectional analysis and exploration of transgender and non-cisgender campaigns and their outcomes.

Variables for campaign outcome present several drawbacks, as discussed above. Without adequate measures for the severity of need across campaigns or for the number of views each campaign page received, it is difficult to develop a single robust outcome variable. Ideally, future researchers would be able to access an indicator of how many page views each campaign gets, so as to then compare various outcome measures against this. The analysis presented here does not focus extensively on other metrics of campaign engagement aside from donation – including shares, likes, and comments. While these were not found to be powerful dependent variables, it is possible that by leaving these out, the analysis has not adequately captured the engagement of broader audiences who do not donate but do share, like, or comment on campaigns. This impact might be more acute for campaigns appealing to networks with less financial capital and would be worth examining more closely in future studies.

Social media platforms are often designed and managed with little attention to how they may create or exacerbate disparities, and attention to disparities may be occluded by platform caompanies' insistence that they are providing a public good. Given that MCF platforms are directly impacting patients' ability to access and afford healthcare, it is essential that these dynamics be further examined. Given that much of the effect here may be due to social "crowd" dynamics, it is likely that any platform which enables healthcare access through crowdfunding is likely to create and exacerbate, rather than address, inequities and precarity. This is similarly true of platforms such as Uber, Airbnb, and Mechanical Turk – key players in a "gig economy" that contribute to economic precarity for user-employees, exacerbate broader social problems, and are difficult to regulate.(82,83) Thus, policymakers should aim to both address disparities within MCF and also (and more importantly) ensure broader healthcare entitlements and social safety net systems that would ease Americans' reliance on MCF in the first place. A first step towards the former effort would be to encourage public access to MCF data and better transparency from crowdfunding companies. To work toward the latter, it is necessary to recognize that crowdfunding is wholly at odds with, and will never be a replacement for, a rights-based system of care which enables all people, regardless of identity, to access necessary healthcare.

# Acknowledgements

This research project was supported by a Royalty Research Fund grant through the University of Washington (UW). Zhihang Dong's participation in the project was supported by the UW Department of Statistics. We are especially thankful to: Ethan Abeles for facilitating data capture; Jessica Cole for early data collection and coding; Fatima Mirza and Anjelica Mendoza for coding and thoughtfully commenting on race data from campaigns; Christine Hahn for her assistance on the broader project from which this research emerges; and Anjum Hajat and Jin-Kyu Jung, whose insights on disparity in crowdfunding informed this analysis.

# Supporting Information

**S1 Fig. Pairwise correlation of social engagement variables and average donation amount.** Pairwise correlation of key co-variates of social engagement by crowdfunders (updates, photos, videos) and with their campaigns (hearts, comments) with average donation amount to the campaign.

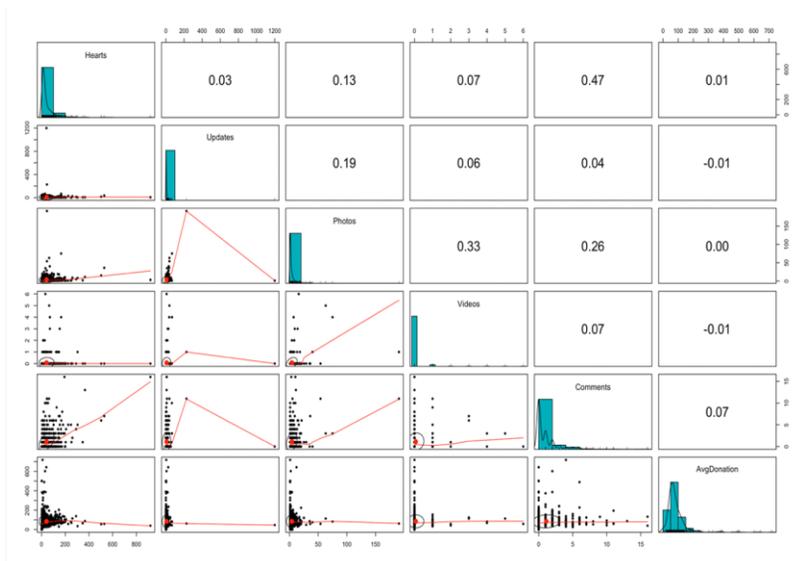

**S2 Fig.**
**Pairwise correlation of social engagement variables and number of donations.** Pairwise correlation of key co-variates of social engagement by crowdfunders (updates, photos, videos) and with their campaigns (hearts, comments) with number of donations to the campaign.

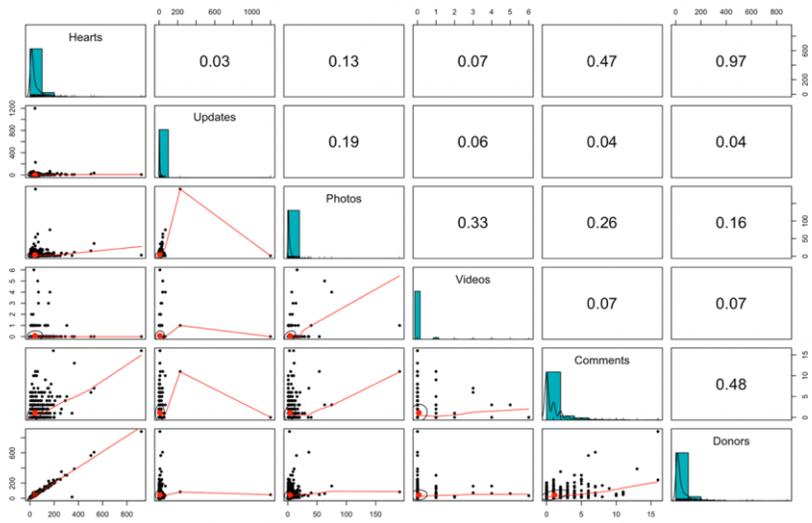